\documentclass[a4paper]{article}
\usepackage[comma,numbers,square,sort&compress]{natbib}
\renewcommand{\top}{{\mathrm{T}}}
\newtheorem{thm}{Theorem}[section]
\newtheorem{lem}[thm]{Lemma}
\newtheorem{cor}[thm]{Corollary}
\newtheorem{rem}[thm]{Remark}
\newtheorem{ass}{Assumption}
\newtheorem{defn}[thm]{Definition}
\usepackage{graphicx,amsmath,amssymb}
\usepackage{subfigure,pgf,tikz,pst-node}
\usetikzlibrary{arrows,shapes,automata,positioning,fit}

\renewcommand{\top}{{\mathrm{T}}}

\begin{document}

\title{Output Average Consensus Over Heterogeneous Multi-Agent Systems via Two-Level Approach}

\author{Yutao Tang \footnote{Yutao Tang is with the School of Automation, Beijing University of Posts and Telecommunications, Beijing 100876. P.\,R. China. Email: yttang@bupt.edu.cn}}

\date{}

\maketitle

{\noindent\bf Abstract}: In this paper, a novel two-level framework was proposed and applied to solve the output average consensus problem over heterogeneous multi-agent systems. This approach is mainly based on the recent technique of system abstraction. For given multi-agent systems, we first constructed their abstractions as the upper level and solved their average consensus problem by leveraging well-known results for single integrators. Then the control protocols for physical agents in the lower level were synthesized in a hierarchical way by embedding the designed law for abstractions into an interface between two levels. In this way, the complexity coming from heterogeneous dynamics of agents is totally decoupled from that of the coordination task and the communication topologies. An example was given to show its effectiveness.

{\noindent \bf Keywords}: average consensus, system abstraction, two-level control,  heterogeneous multi-agent system

\section{Introduction}

As a fundamental problem of multi-agent systems, consensus has been widely investigated due to its numerous applications such as cooperative control of unmanned aerial vehicles, communication among sensor networks, and formation of mobile robots \cite{olfati2004consensus, ren2008distributed, shi2013robust, wang2014consensus, tang2015output}.

While consensus only requires the agreement on some common signal, average consensus (i.e., consensus on the average of some individual signals) of multi-agent agents has been shown as an inevitable part of the solution for more complex problems in several applications, including distributed filtering \cite{carli2008distributed} and multi-robot flocking \cite{tanner2007flocking}. Note that an extra condition has to be satisfied in average consensus, relating the limiting behavior of the whole system to the initial states, this problem is certainly more challenging especially when we expect average consensus of all outputs in a heterogeneous multi-agent network.

The main difficulty hurdling the output average consensus analysis over general multi-agent systems lies in the complexities due to the couplings between high-order heterogeneous dynamic and average task.  In fact, existing results are relatively few, and most publications emphasize on only simple agents (e.g.~single or double integrators) under a fixed graph. For example, a local averaging protocol for single integrators was employed in \cite{olfati2004consensus} from a control viewpoint to solve this problem under a strongly connected and balanced topology.  Fast linear iterations were also proposed in \cite{xiao2004fast} to solve the distributed averaging problem in a discrete-time version. A dynamic average consensus rule was proposed in \cite{kia2014dynamic} such that each agent can track the average of their dynamic inputs with some steady-state error. A similar design was also provided for integrator-type agents under general graphs in \cite{cai2012average}.  However, to our knowledge, there is no general output average consensus conclusion over high-order multi-agent systems even under fixed undirected graphs except \cite{rezaee2015average}. The authors in \cite{rezaee2015average} assumed that each agent shares its initial position and proposed an integral control rule to drive several identical high-order integrators to achieve an output average consensus. Nevertheless, the solvability of output average consensus over general (heterogeneous) linear multi-agent systems is still unclear.

Based on aforementioned observations, an intuitive idea to achieve average consensus over general multi-agent systems is to decouple those complexities brought by high-order dynamics and the average consensus task.  If so, we can separately solve two simpler subproblems instead of the original difficult one. The aim of this paper is to show that the {\it abstraction} technique is possibly the right tool to achieve this goal.

Abstraction, frequently used in computer science \cite{hristu2005handbook} , has been shown as an effective approach to reduce the dynamic complexities of control systems. The authors in \cite{girard2009hierarchical} proved that a given linear system, satisfying some linear matrix inequalities, can be abstracted into its $\Pi$-related system with a lower dimension while keeping the difference between their output trajectories within a computable bound. A hierarchical distributed control approach was later proposed for coordination of general linear multi-agent systems in \cite{tang2013acta}. Based on a simpler abstraction and proper interfaces under system-inclusion assumptions, the coordination protocol was synthesized by some embedding techniques according to difference tasks.

By formulating a two-level control framework via abstraction, we aim to solve the output average consensus problem over general linear multi-agent systems. The contribution of the work is at least twofold. Firstly, comparing with the output consensus or synchronization results in \cite{ren2008distributed, ma2010necessary, scardovi2009synchronization}, we emphasized on the more challenging output average consensus problem. By this novel hierarchical control approach, output average consensus was achieved for a group of general linear agents under mild connectivity conditions, while most existing average consensus analysis and designs were only for special types of dynamics, e.g. single integrators \cite{olfati2004consensus}. Secondly, comparing with the abstraction and hierarchical design for a single system in \cite{girard2009hierarchical}, we provided a more precise characterization for linear abstraction problem and extended it to a distributed version for networked systems. Furthermore, those agents considered here are heterogeneous, while only the homogeneous case was considered in \cite{tang2013acta}.

The rest of this paper is organized as follows. Problem statement is presented in Section \ref{sec:formulation}. After introducing some basic results of system abstraction in Section \ref{sec:abs}, we provide a two-level control scheme with main results in Section \ref{sec:main}. Finally, simulation examples and concluding remarks are presented at the end.

\textbf{Notations}: let $\mathbb{R}^n$ be the $n$-dimensional Euclidean space, $\mathbb{R}^{n\times m}$ be the set of $n\times m$ real matrices. $\text{diag}\{b_1,{\dots},b_n\}$ denotes an $n\times n$ diagonal matrix with diagonal elements $b_i\; (i=1,{\dots},n)$; $\text{col}(a_1,{\dots},a_n) = [a_1^{\top},\, {\dots},\,a_n^{\top}]^{\top}$ for any column vectors $a_i\; (i=1,{\dots},n)$. For any vector $y\in \mathrm{R}^n$, $\mathrm{ Ave}(y)\triangleq\frac{\sum_{i=1}^n y_i}{n}$. A continuous function $\alpha\colon[0,\, a)\to [0,\, \infty)$ belongs to class $\mathcal{K}$ if it is strictly increasing and $\alpha(0)=0$; It belongs to class $\mathcal{K}_\infty$ if it belongs to class $\mathcal{K}$ with $a=\infty$ and $\lim_{s\to \infty}\alpha(s)\to\infty$.

\section{Problem Statement}\label{sec:formulation}

Consider a group of (non-identical) linear agents of the form
\begin{align}\label{sys:agent}
\begin{cases}
\dot{x}_i=A_i x_i+ b_i u_i\\
y_i=c_ix_i, \quad i=1,{\dots},N
\end{cases}
\end{align}
where $x_i \in \mathbb{R}^{n_{x_i}},\; u_i\in \mathbb{R}^{n_{io}}, \,y_i\in\mathbb{R}^{n_{io}}$ are its state, input and output, respectively. $A_i,\, b_i,\,c_i$ are matrices with proper dimensions. We assume the pair $(c_i,\, A_i,\, b_i)$ is minimal and has no transmission zero at the origin of complex plane. Apparently, it includes minimum-phase linear systems and of course high-order integrators as its special cases. For simplicity, we take $n_{io}=1$. The following arguments can be extended to multiple-input and multiple-output cases without difficulties.

Associated with these multi-agent systems, a digraph $\mathcal{G}=\{\mathcal{V},\,\mathcal{E},\, \mathcal{A}\}$ is defined with node set $\mathcal{N}=\{1,\,...,\,N\}$ to describe the communication topology (see \cite{mesbahi2010graph} for details). If agent $i$ can get access to the information of agent $j$, there exists a directed edge in $\mathcal{G}$ with weight $a_{ij}$.  Define the neighbor set of agent $i$ as $\mathcal{N}^0_i=\{j: (j,i)\in \mathcal {E}\}$ for $i=1,\,...,\,N$ and $\mathcal{N}_i=\mathcal{N}^0_i\cup \{i\}$. To achieve a coordination of those agents, this graph should be connected to some degree. The following assumption has been widely in multi-agent systems \cite{olfati2004consensus, ren2008distributed}.

\begin{ass} \label{ass:graph}
	The graph $\mathcal{G}$ is undirected and connected.
\end{ass}

Invoking the topology of these multi-agent systems, we consider distributed control laws of the following form.
\begin{align}\label{ctr:output}
\begin{split}
u_i&={\phi}_i(\xi_j,\, x_j,\, j\in \mathcal{N}_i)\\
\dot{\xi}_i&={\psi}_i(\xi_j,\, x_j,\, j\in \mathcal{N}_i)
\end{split}
\end{align}
where $\xi_i\in \mathbb{R}^{n_{\xi_i}}$ with a nonnegative integer $n_{\xi_i}$ and smooth functions ${\phi}_i$,\,${\psi}_i$ to be designed later. When $n_{\xi_i}=0$, the control law for agent $i$ is a static one. It becomes an output feedback control when only $y_j$ (not $x_j$) ($j\in \mathcal{N}_i$) is used.

Next, we formulate the output average consensus problem over these multi-agent systems as follows. {\it Given multi-agent systems composed of \eqref{sys:agent} and a communication graph $\mathcal{G}$, find a distributed control law $u_i$ of the form \eqref{ctr:output}, such that
	\begin{equation}\label{errori}
	\lim_{t\to \infty}e_i(t)=0, \,i=1,\,{\dots},\,N,
	\end{equation}
	where $e_i\triangleq y_i-\mathrm{ Ave}(y(0)).$
}

\begin{rem}\label{rem:def}
	Instead of merely reaching a consensus, an extra condition, $y_i(\infty)=\mathrm{ Ave}(y(0))$, has to be satisfied which relates the limiting behavior of those multi-agent systems to their initial states. In contrast to many publications on output consensus among several classes of multi-agent systems (e.g.~\cite{ hong2006tracking, ma2010necessary}), few results are obtained on average consensus for agents. Moveover, existing average consensus results are only derived for some special multi-agent systems, e.g. integrators \cite{olfati2004consensus}, while heterogeneous multi-agent systems with general linear dynamics are considered here, which makes this problem much more challenging.
\end{rem}

In the following sections, we first give a brief introduction to abstraction techniques, then present a novel hierarchical control approach for our multi-agent system to solve its output average consensus problem.

\section{Abstraction and Simulation Function}\label{sec:abs}
In this section, preliminary knowledge on abstraction is given for the following analysis. For the sake of consistency, we emphasize on single-input and single-output linear system as follows.
\begin{align}\label{sys:concrete}
S:\begin{cases}
\dot{x}=Ax+bu\\
y=cx
\end{cases}
\end{align}
where $x \in \mathbb{R}^n,\, u \in \mathbb{R},\,y(t)\in \mathbb{R}$. Suppose $(c,\,A,\,b)$ is minimal. In the following, we refer to (\ref{sys:concrete}) as the concrete system to be controlled in practice.

We adopt the simulation-based abstraction formulation \cite{girard2009hierarchical} for our design and suppose the abstraction of (\ref{sys:concrete}) is also linear of the form:
\begin{equation}\label{sys:abstraction}
S':\begin{cases}\dot{z}=Fz+gv\\
w=hz\end{cases}
\end{equation}
where $z \in \mathbb{R}^m, v \in \mathbb{R}, w\in \mathbb{R}$.  Typically, the dimension of system $\eqref{sys:abstraction}$ is much smaller than that of the concrete system, i.e., $m \ll n$.

Throughout the paper $S$ and $S'$ are supposed to be forward complete, namely, for every initial condition and every measurable locally essentially bounded input signal, the solution is defined for all $t\geq 0$, i.e., the maximal interval of existence is $T_{\max}=+\infty$. Inspired by \cite{girard2009hierarchical} and \cite{tang2013acta}, we introduce simulation functions as follows to quantify the approximation relationship between concrete system and its abstraction.
\begin{defn}\label{def:simu}
	Let $V\colon \mathbb{R}^m \times \mathbb{R}^n\rightarrow \mathbb{R}^+$ be a
	smooth function with $V(0,0)=0$ and $u_v:\mathbb{R}\times
	\mathbb{R}^m\times \mathbb{R}^n \rightarrow \mathbb{R}$ be a
	continuous function. $V$ is a simulation function of $S'$ by $S$ and
	$u_v$ is an associated interface if there exist two class $\mathcal{K}$
	functions $\gamma$ and $\alpha$ such that, for all
	$(z,x)\in \mathbb{R}^m \times \mathbb{R}^n$,
	\begin{equation}\label{eq:sim1}
	V(z,x)\geq (y-z)^2
	\end{equation}
	and for all $v\in \mathbb{R}$ satisfying $\gamma(||v||)<V(z,x)$,
	\begin{align}\label{eq:sim2}
	\frac{\partial{V}}{\partial{z}} (Fz+Gv)+\frac{\partial{V}}{\partial{x}} (Ax+bu_v)<-\alpha(V).
	\end{align}
\end{defn}

In Inequality (\ref{eq:sim2}), we impose a strict decreasing rate than that in \cite{girard2009hierarchical}, which benefits us with more precise characterization of the simulation relationship between those outputs of $S'$ and $S$.

\begin{lem}\label{lem:abs:boudness}
	Let $V$ be a simulation function of $S'$ by $S$ with an associated
	interface $u_v$, $v(t)$ be an admissible input
	of $S'$, $x$,\, $z$ and $w$ satisfying
	\begin{align}
	\begin{split}\label{sys:composite}
	\dot{x}&=Ax+bu_v(z,x,v),\\
	\dot{z}&=Fz+gv,\\
	y&=cx,\quad w=hz.
	\end{split}
	\end{align}
	Then, there exists a class $\mathcal{KL}$ function $\beta(\cdot)$ holding for $t\geq 0$
	\begin{align}\label{eq:lem-bounded-kl}
	(y-w)^2\leq \beta(V(x(0),z(0)),\,t)+\gamma(||v||_{\infty}).
	\end{align}
	Additionally, if $v(t)$ vanishes when $t$ goes to infinity, then,
	\begin{align}\label{eq:lem-bounded-0}
	|w(t)-y(t)|\to 0 \quad(t\to+\infty).
	\end{align}
\end{lem}
\noindent\textbf{Proof}. The proof will follow the traditional technique of comparison functions and is split into two steps.

\noindent\textit{Step 1}: Consider a set $M\triangleq \{(z,x)\colon V(z,x)\leq c_0\}$, where $c_0\triangleq \gamma(||v||_\infty)$. We claim that, if there exists a $t_0$ such that $(z(t_0),x(t_0))\in M$, then $(z(t),x(t))\in M$ for $t>t_0$. We prove it by seeking a contradiction.

Assume there exist some $t>t_0, \epsilon$ and $V(z(t),x(t))>c_0+\epsilon$. Let $\tau=\inf(t>t_0:V(z(t),x(t))>c_0+\epsilon)$, then $V(z(\tau),x(\tau))\geq c_0+\epsilon$, which implies $V(z(\tau),x(\tau))\geq c$. Combining with the definition of $V$, it gives
\begin{align}
\frac{\partial{V}}{\partial{t}}\mid_{t=\tau}<-\alpha(V(z,x)\mid_{t=\tau})<0
\end{align}
Thus, $V(z(t),x(t))>V(z(\tau),x(\tau))$ for some $t$ during $(t_0,\tau)$, which
contradicts to the minimality of $\tau$. Hence the following argument holds as we claimed:
\begin{align}\label{eq:lem-bounded:eq1}
V(z(t),x(t))<\gamma(||v||_\infty), \mbox{ for all } t>t_0.
\end{align}

\noindent\textit{Step 2}: Let $t_1=\inf\{t>0:(z(t),x(t))\in M\}$. It follows from the above
argument $V(z(t),x(t))\in M$ for $t>t_1$, i.e., $V(z(t),x(t))\leq
\gamma(||v||)$. For any $t<t_1$, $(z(t),x(t))\notin M$, that is, $V(z(t),x(t))>c=\gamma(||v||_\infty)$. By the definition of simulation function, we have, for $0\leq t<t_1$,
\begin{align*}
\frac{\partial{V}}{\partial{t}}<-\alpha(V(z(t),x(t))).
\end{align*}
By a stand comparison principle \cite{khalil2002nonlinear},  there exists some class $\mathcal{KL}$ function $\beta$ such that
\begin{align*}\label{eq:lem-bounded:eq2}
\!V(z(t),x(t))<\beta(V(z(0),x(0)),t),\mbox{ for all }t<t_1.
\end{align*}

Combining these inequalities, one can safely conclude
\begin{align*}
(w(t)-y(t))^2<\beta(V(z(0),x(0)),t)+\gamma(||v||_\infty)
\end{align*}
which is exactly the inequality \eqref{eq:lem-bounded-kl}.

Suppose $v(t)$ vanishes as $t$ goes to infinity, then for any given $\epsilon>0$, there exists a large enough $T$ satisfying $|v(t)|\leq \gamma^{-1}(\frac{\epsilon}{2})$ for $t> T$. We can apply the previous arguments to the composite system from a new initial time $t_0=T$, and thus obtain,
\begin{align*}
(w(t)-y(t))^2&<\beta(V(z(T),x(T)),t)+\frac{\epsilon}{2}
\end{align*}

Note that the function $\beta(\cdot,\,\cdot)\in \mathcal{KL}$, there exists a $T_1>0$ such that
\begin{align*}
\beta(V(z(T),x(T)),t)<\frac{\epsilon}{2},\quad \forall t>T_1
\end{align*}

Taking $T_0=\max\{T,\, T_1\}$, one can thus obtain that, for any $t>T_0>0$,
\begin{align*}
(w(t)-y(t))^2&<\epsilon
\end{align*}
That is $w(t)-y(t)\to 0$ as $t\to+\infty$. We thus complete the proof.\hfill\rule{4pt}{8pt}

A similar estimation was established in \cite{girard2009hierarchical}:
\begin{equation*}
|y(t)- w(t)|\leq \max\{V(x(0),z(0)),\,\gamma(||v||_\infty)\}
\end{equation*}
Under Definition \ref{def:simu}, it allows the concrete system to asymptotically simulate its abstraction for some special design of $v$, which is more proper for consensus coordination over complex multi-agent systems.

\section{A Two-Level Design}\label{sec:main}
\begin{figure}
	\centering \includegraphics[width=0.65\textwidth]{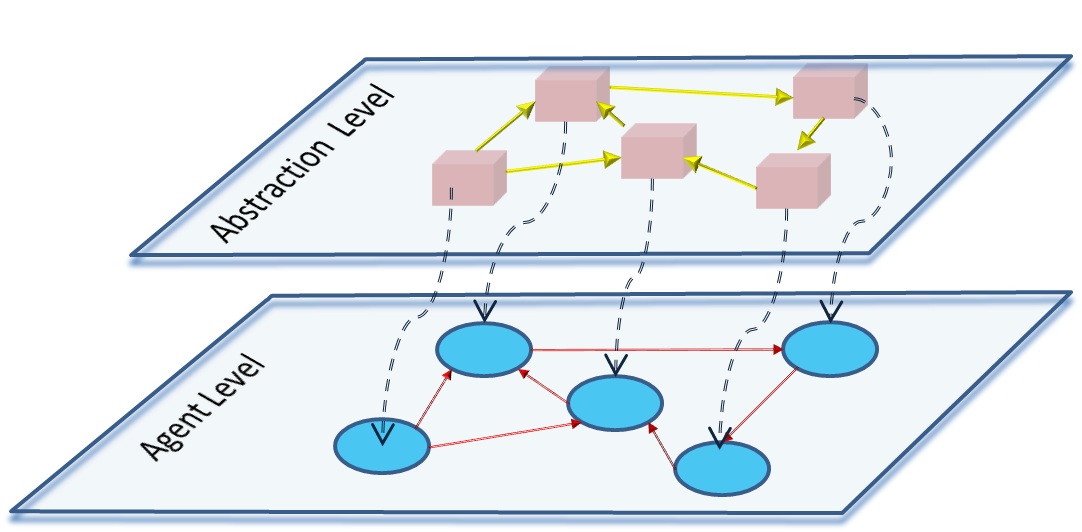}
	\caption{Two-level structure of abstraction-based hierarchical control.}\label{fig:info-flow}
\end{figure}

In this section, we will provide a two-level framework to solve the output average consensus problem over heterogeneous multi-agent systems.  The hierarchical structure and information flow of our abstraction-based distributed control approach is depicted in Figure \ref{fig:info-flow}. Details will be specified in the following subsections.

\subsection{Abstraction Level: Protocol Design}
First, a network of abstractions should be set up as the upper level according to given tasks. The following lemma for linear systems plays a key role in our design.
\begin{lem}\label{lem:ab:integrator}
	If the triple $(c,\, A,\, b)$ is minimal with no zero at the origin, the system \eqref{sys:concrete} takes an integrator $\dot{z}=v,\, w=z$ as its abstraction with an interface of the form $u=k(x-Xz)+Uz+Rv$, where $X,\,U$ satisfy $AX + bU=0,\; cX=1$, $k$ is a matrix such that $A+bk$ is Hurwitz and $R$ is freely selected.
\end{lem}

\noindent\textbf{Proof.} To prove this lemma, we only have to find a simulation function and its associated interface. Since the plant has no zero at the origin, by the transmission zero condition in \cite{huang2004nonlinear}, there exists a pair $(X,\, U)$ satisfying $AX+bU=0,\; cX=1$. Taking a matrix $k$ such that $A+bk$ is Hurwitz and letting $\bar x=x-Xz$ give
\begin{align*}
\dot{\bar x}=Ax+bu-Xv,\quad  y-w=c\bar x
\end{align*}
Next, we take a linear control law $u_v=kx+Uz+Rv$,
\begin{align*}
\dot{\bar x}=(A+bk)\bar x+(bR-X)v.
\end{align*}
One can easily verify the conditions in Definition \ref{def:simu}.

In fact, since $A+bk$ is Hurwitz, there exists a positive definite matrix $P\in \mathbb{R}^{n\times n}$ such that $(A+bk)^{\top}P+P(A+bk)=-I_n$. Take $V=\hat c\bar x^{\top}P\bar x$, where $\hat c\geq \frac{||c||^2}{\lambda_{\min}(P)}$.  Apparently, $V\geq||c(x-Xz)||^2=(y-z)^2$.
\begin{align*}
\dot{V}&=2\hat c\bar x^{\top}[(A+bk)\bar x+(bR-X)v]\\
&\leq -\hat c\bar x^{\top}\bar x+2\hat c\bar x^{\top}P(bR-X)v\\
&\leq -\frac{V}{2\lambda_{\max}(P)}-(\frac{V}{4\lambda_{\max}}-4c||P(bR-X)||^2v^2)
\end{align*}
Let $\alpha(s)=\frac{1}{2\lambda_{\max}(P)}s$ and $$\gamma(s)=16\lambda_{\max}(P)||P(bR-X)||^2 s^2,$$ the inequality (\ref{eq:sim2}) holds. Thus $V$ and $u_v$ are indeed the simulation function and associated interface between $(c,\,A,\,b)$ and the integrator under Definition \ref{def:simu}. \hfill\rule{4pt}{8pt}

By Lemma \ref{lem:ab:integrator}, we select the abstraction of agent $i$ to achieve average consensus in the upper level as follows:
\begin{align}\label{sys:abs-i}
\begin{split}
\dot{z}_i=v_i, \quad w_i=z_i
\end{split}
\end{align}
Then, proper protocols to complete the average consensus goal in the upper level for those abstractions should be proposed, which has been well-studied in literature \cite{olfati2004consensus}.

A neighbor-based protocol is used here for average consensus of abstractions:
\begin{align}\label{ctr:abs}
v_i=\sum_{j=1}^N a_{ij}(z_j-z_i).
\end{align}

\begin{rem}\label{rem:weighted}
	Although we only consider output average consensus for agents of the form \eqref{sys:agent}, it is remarkable that output weighted consensus problem on $y^*_{w}\triangleq\frac{\sum_{i=1}^N w_i y_i(0) }{\sum_{i=1}^N w_i}$ with $w_i>0$ can also be solved when weighted consensus is achieved in the upper level. In fact, letting $v_i=\frac{1}{w_i}\sum_{j=1}^N a_{ij}(z_j-z_i)$ and recalling Corollary 3 in \cite{olfati2004consensus}, we can solve it easily by following the same procedure.
\end{rem}

\subsection{Agent Level: Interface Embedding}

After designing controllers for the abstraction level, we synthesis the control protocol for \eqref{sys:agent} to solve our original multi-agent systems.

Note that, the abstraction is locally set up as a virtual reference, it is reasonable to choose its initial conditions and use its state and input in our control as well.  Thus the distributed coordination is transformed into $N$ decentralized interface constructing and reference embedding problems.

Combined with Lemma \ref{lem:ab:integrator}, a state-feedback interface candidate would be $u_i=k_i(x_i-X_iz_i)+U_iz_i+R_iv_i$. by embedding \eqref{ctr:abs} into this interface, a control candidate to achieve output average consensus of our multi-agent systems is
\begin{align}\label{ctr:state}
\begin{split}
u_i&=k_i(x_i-X_iz_i)+U_iz_i+R_i\sum_{j=1}^N a_{ij}(z_j-z_i),\\
\dot{z}_i&=\sum_{j=1}^N a_{ij}(z_j-z_i), \quad z_i(0)=y_i(0).
\end{split}
\end{align}

When only output information $y_i$ is available, it is possible to provide a dynamic output-feedback interface as follows
\begin{align}
u_i&=k_i(\xi_i-X_iz_i)+U_iz_i+R_i\sum_{j=1}^N a_{ij}(z_j-z_i),\nonumber\\
\dot{\xi}_i&=A_i\xi_i+b_iu_i-l_i(y_i-c_i\xi_i),\nonumber\\
\dot{z}_i&=\sum_{j=1}^N a_{ij}(z_j-z_i), \quad z_i(0)=y_i(0)\label{ctr:output}
\end{align}
where $l_i$ is a chosen matrix such that $A_i+l_ic_i$ Hurwitz.

It is interesting to remark that an initialization $z_i(0)=y_i(0)$ is employed in our design, which can be taken as an information-sharing mechanism to achieve a cooperative task. Similar ideas have been used in many publications for integrators, e.g.~\cite{kia2014dynamic}  and  \cite{rezaee2015average}.

\subsection{Performance Evaluation and Convergence}
In this subsection, we shall prove the solvability of output average consensus problem over these heterogeneous multi-agent systems under proposed controls \eqref{ctr:state} and \eqref{ctr:output}.

It is time to give our first main theorem.
\begin{thm}\label{thm:state}
	Under Assumption \ref{ass:graph}, the output average consensus of multi-agent systems \eqref{sys:agent} can be achieved by a state-feedback control law of the form \eqref{ctr:state}.
\end{thm}

\noindent\textbf{Proof.} To prove this theorem, we rewrite the composite system as follows.
\begin{align*}
\dot{x}_i&=A_ix_i+b_i[k_i(x_i-X_iz_i)+U_iz_i+R_iv_i]\\
\dot{z}_i&=\sum_{j=1}^N a_{ij}(z_j-z_i), \quad z_i(0)=y_i(0)
\end{align*}
where $v_i=\sum_{j=1}^N a_{ij}(z_j-z_i)$. It is of the cascaded form while $x_i$-subsystem is the driven one. Letting $z=\mbox{col}(z_1,\,\cdots,\,z_N)$ gives
\begin{align*}
\dot{z}&=-Lz, \quad z(0)=y(0)
\end{align*}
By Theorem 5 in \cite{olfati2004consensus}, it solves the average consensus of $z$, that is, $\lim\limits_{t\to \infty}z_i(t)=\mathrm{ Ave}(z(0))=\mathrm{ Ave}(y(0))$.

Apparently, it happens that $\lim\limits_{t\to\infty}v_i=0$. By Lemma \ref{lem:ab:integrator}, we have
\begin{align*}
|e_i(t)|\leq |y_i(t)-w_i(t)|+|z_i(t)-\mathrm{ Ave}(y(0))|\to 0
\end{align*}
as $t\to +\infty$ and thus complete the proof. \hfill\rule{4pt}{8pt}

This result can be extended to a class of digraphs as follows.
\begin{cor}\label{cor:state}
	For given multi-agent systems composed of \eqref{sys:agent} with a fixed digraph topology $\mathcal{G}$. The output average consensus problem is solvable if and only if $\mathcal{G}$ is strongly connected and balanced.
\end{cor}
The proof is not hard, since the average consensus in the upper level under \eqref{ctr:abs} is solved if and only if $\mathcal{G}$ is strongly connected and balanced by Theorems 3 and 5 in \cite{olfati2004consensus}

There might be some cases when only $y_i$ is available through measurement. We then present the second main theorem for these multi-agent systems under output feedback law \eqref{ctr:output}.
\begin{thm}\label{thm:output}
	Under Assumption \ref{ass:graph}, the output average consensus of multi-agent systems composed of \eqref{sys:agent} can be solved by a dynamic output-feedback control law of the form \eqref{ctr:output}.
\end{thm}

\noindent\textbf{Proof.} As that in Theorem \ref{thm:state}, the composite system under \eqref{ctr:output} is as follows.
\begin{align*}
\dot{x}_i&=A_ix_i+b_i[k_i(x_i-X_iz_i)+U_iz_i+R_iv_i]+b_ik_i(\xi_i-x_i)\\
\dot{\xi}_i&=A_i\xi_i+b_iu_i-l_i(y_i-c_i\xi_i),\\
\dot{z}_i&=\sum_{j=1}^N a_{ij}(z_j-z_i), \quad z_i(0)=y_i(0)
\end{align*}

Letting $z=\mbox{col}(z_1,\,\cdots,\,z_N)$, $\bar \xi_i=\xi_i-x_i$ and $\bar x_i=x_i-X_iz_i$ gives
\begin{align*}
\dot{\bar x}_i&=(A_i+b_ik_i)\bar x_i+(b_iR_i-X_i)v_i+b_ik_i\bar \xi_i\\
\dot{\bar \xi}_i&=(A_i+l_ic_i)\bar \xi_i,\\
\dot{z}&=-Lz, \quad z(0)=y(0).
\end{align*}
Since $A_i+b_ik_i$ and $A_i+l_ic_i$ are both Hurwitz by selections and $\lim\limits_{t\to\infty}v_i=0$, we thus obtain
$|y_i(t)-z_i(t)|=|c_i\bar x_i|\to 0$ as $t$ goes to infinity by Theorem 9.1 in \cite{khalil2002nonlinear}. Recalling $\lim\limits_{t\to \infty}z_i(t)=\mathrm{Ave}(z(0))=\mathrm{Ave}(y(0))$ in Theorem \ref{thm:state}, we thus complete the proof. \hfill\rule{4pt}{8pt}

For agents of special dynamics, static output feedback interfaces might be constructed, which brings reduced-order output feedback controls to achieve output average consensus of those heterogeneous multi-agent systems.

\begin{thm}\label{thm:static-output}
	Consider the multi-agent systems composed of \eqref{sys:agent} with a fixed topology satisfying Assumption \ref{ass:graph}. Suppose for any $i$, there exists a constant $\nu_i\neq 0$ satisfying
	\begin{align}\label{eq:static:eq1}
	b_i^{\top}P_i=\nu_i c_i
	\end{align}
	where $P_i$ is positive definite and satisfies the following linear matrix inequality
	\begin{align}\label{eq:static:eq2}
	A_i^{\top}P_i+P_iA_i<2P_ib_ib_i^{\top}P_i.
	\end{align}
	Then, a reduced-order output feedback of the form
	\begin{align}
	\begin{split}
	u_i&=k_{i}(y_i-z_i)+U_{i}z_i+R_{i}\sum_{j=1}^N a_{ij}(z_j-z_i),\\
	\dot{z}_i&=\sum_{j=1}^N a_{ij}(z_j-z_i), \quad z_i(0)=y_i(0)
	\end{split}
	\end{align}
	can be employed to solve the average consensus problem for these multi-agent systems.
\end{thm}

\noindent\textbf{Proof}. Following the procedure of our hierarchical control approach, we have $y_i-z_i=c_i(x_i-X_iz_i)$.
The composite system is then as follows.
\begin{align*}
\dot{x}_i&=A_ix_i+b_i[k_ic_i(x_i-X_iz_i)+U_iz_i+R_iv_i]\\
\dot{z}_i&=\sum_{j=1}^N a_{ij}(z_j-z_i), \quad z_i(0)=y_i(0).
\end{align*}
Again, letting $z=\mbox{col}(z_1,\,\cdots,\,z_N)$ and $\bar x_i=x_i-X_iz_i$ gives
\begin{align*}
\dot{\bar x}_i&=(A_i+b_ik_ic_i)\bar x_i+(b_iR_i-X_i)v_i\\
\dot{z}&=-Lz, \quad z(0)=y(0).
\end{align*}
We claim that there exists a matrix $k_i$ such that $A_i+b_ik_ic_i$ is Hurwitz. If so, the proof can be completed without difficulties by following the procedure in Theorem \ref{thm:state}.

In fact, from \eqref{eq:static:eq1}, we have $b_i^{\top}P_i=\nu_i c_i$. Letting $k_i=-\hat \lambda_i \nu_i$ gives
\begin{align*}
M_i&\triangleq(A_i+b_ik_ic_i)^{\top}P_i+P_i(A_i+b_ik_ic_i)\\
&=(A_i-\hat \lambda_i b_ib_i^{\top}P_i)^{\top}P_i+P_i(A_i-\hat \lambda_i b_ib_i^{\top}P_i)\\
&= A_i^{\top}P_i+P_iA_i-2\hat \lambda_i P_ib_ib_i^{\top}P_i.
\end{align*}
Recalling \eqref{eq:static:eq2} and letting $\hat \lambda_i>1$ gives
\begin{align*}
M_i\triangleq(A_i+b_ik_ic_i)^{\top}P_i+P_i(A_i+b_ik_ic_i)<0
\end{align*}
Since $P_i$ is positive definite, $A_i+b_ik_ic_i$ is Hurwitz by Theorem 4.6 in \cite{khalil2002nonlinear}. The proof is complete. \hfill\rule{4pt}{8pt}

This result can be extended to multiple input and multiple-output systems by assuming
\begin{align}\label{eq:static:eq2}
	\text{Rank}(c_i)=\text{Rank}\begin{pmatrix}
	c_i\\
	b_i^{\top}P_i
	\end{pmatrix}.
\end{align}

\begin{rem}\label{rem:switch}
	Although we only consider multi-agent systems under a fixed topology, these conclusions still hold when the communication topology varies and switches among a finite number of strongly connected and balanced graphs by employing common Lyapunov techniques as that in \cite{olfati2004consensus}.
\end{rem}

\begin{rem}\label{rem:heterogeneous}
	In contrast to many existing average consensus results on single integrators \cite{cai2012average, olfati2004consensus, ren2008distributed}, this abstraction-based framework can handle heterogeneous general multi-agent systems with possible general graphs by a two-level design. In fact, through such an abstraction process, the heterogeneous linear multi-agent agents are homogenized and further converted to well-studied integrator-type systems. That is, the complexity from dynamics is decoupled from that of the average consensus goal and communication topology. In this sense, this two-level approach might provide a promising framework to handle multi-agent systems with both complex dynamics and sophisticated tasks.
\end{rem}

\section{Simulations}\label{sec:simu}
For illustrations, we present an example and consider four agents with following system matrices
\begin{align*}
A_1=1,\; b_1=1,\; c_1=1,
\end{align*}
\begin{align*}
A_2=\begin{bmatrix}
0&1\\-1&0
\end{bmatrix},\; b_2=\mbox{col}(0,1),\; c_2=[1,\;0],
\end{align*}
\begin{align*}
A_3=\begin{bmatrix}
0&1&0\\
-1&0&1\\
2&0&1\\
\end{bmatrix},\; b_3=\mbox{col}(0,1,1),\; c_3=[0,\;1,\;0].
\end{align*}
\begin{align*}
A_4=\begin{bmatrix}
0&1\\0&0
\end{bmatrix},\; b_4=\mbox{col}(0,1),\; c_4=[1,\;0].
\end{align*}

\begin{figure}
	\centering
	\subfigure[The graph $\mathcal{G}_1$]
	{
		\begin{tikzpicture}[shorten >=1pt, node distance=1 cm, >=stealth',
		every state/.style ={circle, minimum width=0.2cm, minimum height=0.2cm}, auto]
		\node[align=center,state](node1) {1};
		\node[align=center,state](node2)[right of=node1]{2};
		\node[align=center,state](node3)[right of=node2]{3};
		\node[align=center,state](node4)[right of=node3]{4};
		\path[->]   (node1) edge [bend left] (node2)
		(node2) edge [bend left] (node3)
		(node3) edge [bend left] (node4)
		(node4) edge [bend left] (node3)
		(node3) edge [bend left] (node2)
		(node2) edge [bend left] (node1)
		;
		\end{tikzpicture}
	}\quad
	\subfigure[The graph $\mathcal{G}_2$]
	{
		\begin{tikzpicture}[shorten >=1pt, node distance=1 cm, >=stealth',
		every state/.style ={circle, minimum width=0.2cm, minimum height=0.2cm}, auto]
		\node[align=center,state](node1) {1};
		\node[align=center,state](node2)[right of=node1]{2};
		\node[align=center,state](node3)[right of=node2]{3};
		\node[align=center,state](node4)[right of=node3]{4};
		\path[->]   (node1) edge [bend left] (node4)
		(node4) edge  (node3)
		(node3) edge (node2)
		(node2) edge (node1)
		;
		\end{tikzpicture}
	}
	\caption{The communication graphs.}\label{fig:graph}
\end{figure}
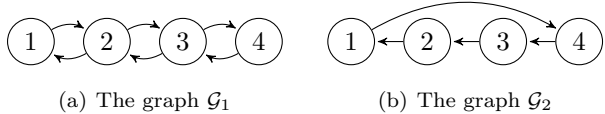

We assume the topology is switching between two strongly connected and balanced graphs $\mathcal{G}_i$ $(i=1,2)$ described by Figure \ref{fig:graph} with unity weights. The switchings are periodically carried out in the following order $\{\mathcal{G}_1, \mathcal{G}_2, \mathcal{G}_1, \mathcal{G}_2, \cdots\}$ with a period $T=5$.

\begin{figure}
	\centering
	\subfigure[]{
		\includegraphics[width=0.42\textwidth]{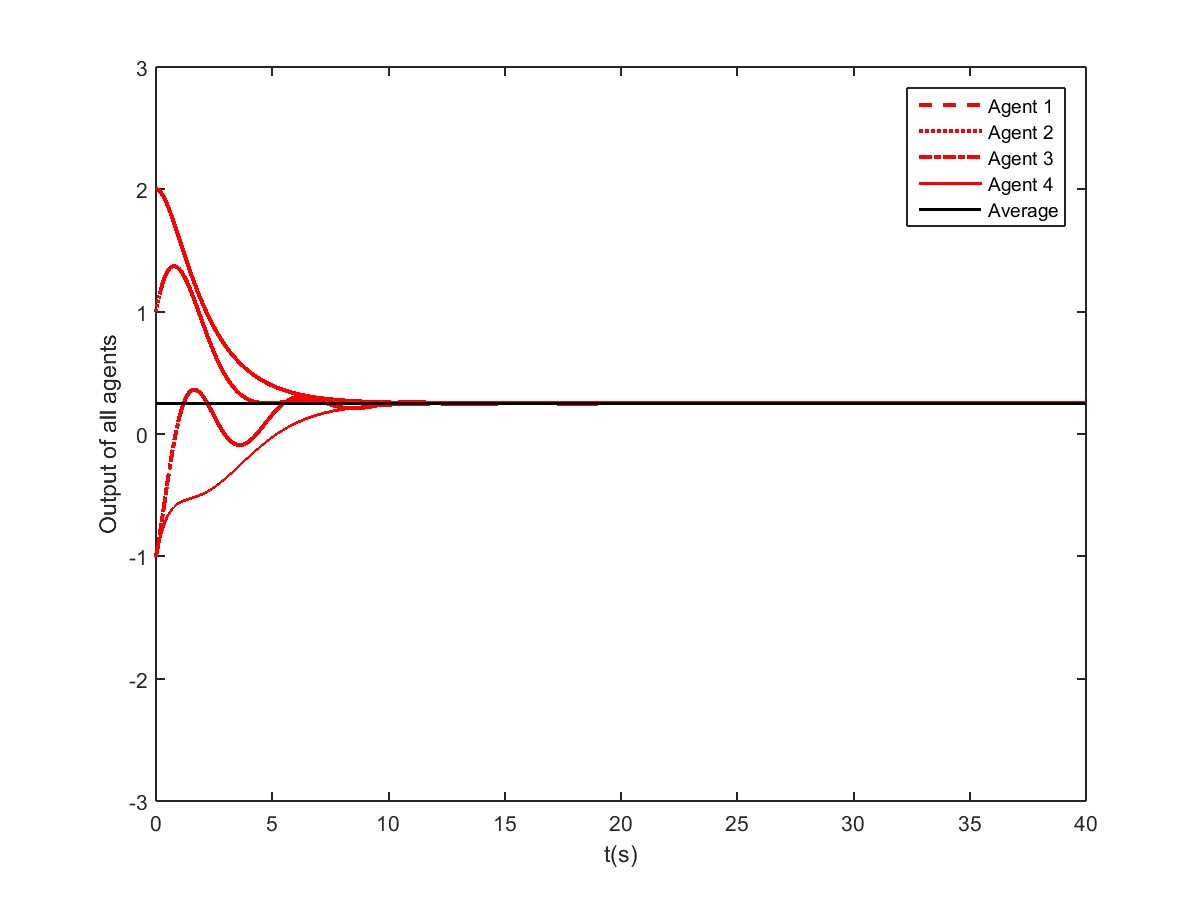}
	}
	\subfigure[]{
		\includegraphics[width=0.42\textwidth]{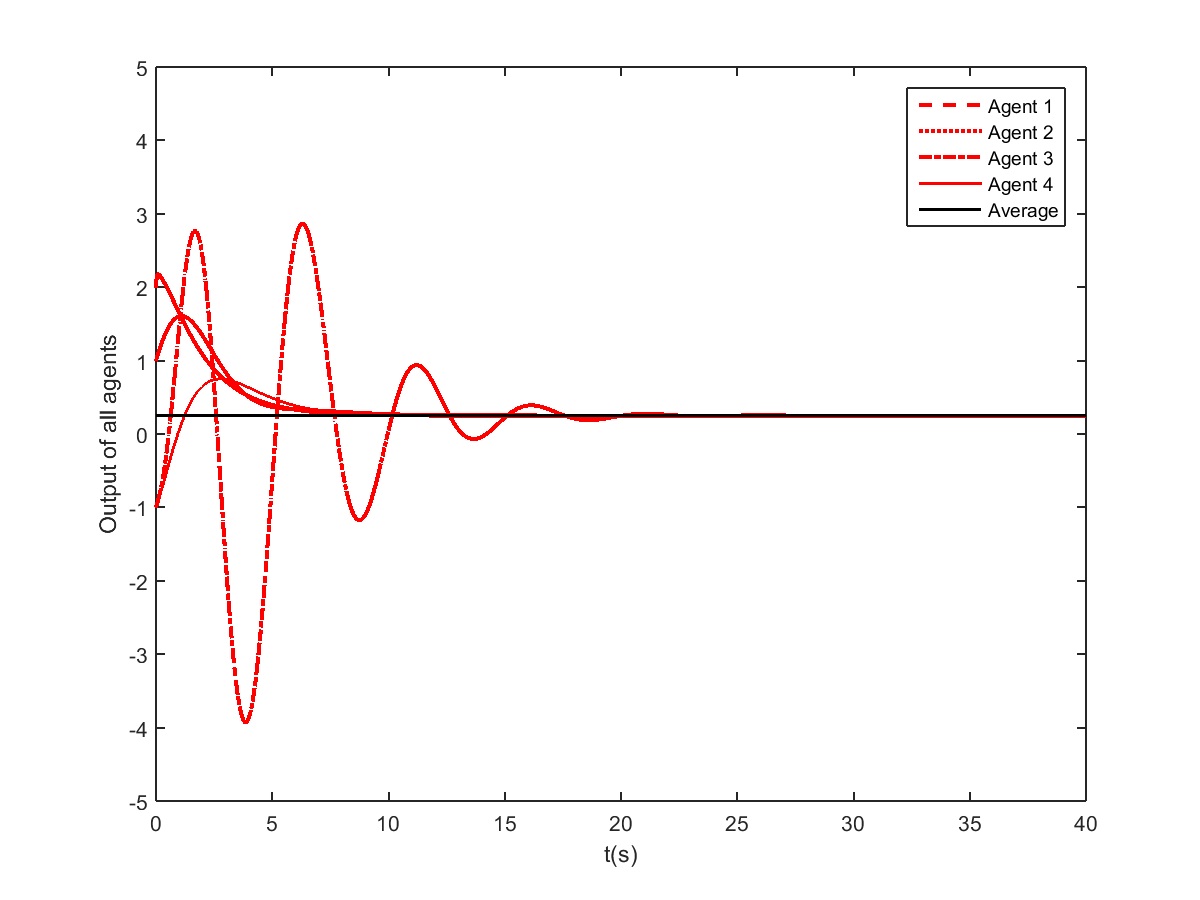}
	}
	\caption{Profiles of all outputs under the controllers \eqref{ctr:state} and \eqref{ctr:output}.}\label{fig:simu}
\end{figure}

It can be verified the systems are minimal with no transmission zero at the origin. By Remark \ref{rem:switch} and some further manipulations, the output average consensus problem of them can be solved by the controllers \eqref{ctr:state} and \eqref{ctr:output}.  The effectiveness is also verified by the simulation results with selected gain matrices depicted in Figure \ref{fig:simu}.

\section{Conclusions}\label{sec:conclusion}
A  two-level approach was proposed to solve the output average consensus problem over heterogeneous multi-agent systems. In conjunction with abstraction techniques, two different protocols were derived for those agents. Future work includes extensions to more general graphs and nonlinear cases.

\bibliographystyle{ieeetr}

\end{document}